\pdfoutput=1
\documentclass[conference]{IEEEtran}
\IEEEoverridecommandlockouts

\usepackage{cite}
\usepackage{amsmath,amssymb}
\usepackage{graphicx}
\usepackage{xcolor}
\usepackage{booktabs}
\usepackage{tabularx}
\usepackage{array}
\newcolumntype{Y}{>{\raggedright\arraybackslash}X}
\newcolumntype{L}[1]{>{\raggedright\arraybackslash}p{#1}}
\usepackage{tikz}
\usepackage{pgfplots}
\pgfplotsset{compat=1.18}
\usepgfplotslibrary{groupplots}
\usepackage{url}
\usepackage[hidelinks]{hyperref}

\IfFileExists{generated/macros.tex}{\newcommand{\PowerOffOutage}{1.34}  %
\newcommand{\BlockOutageMin}{3.19}  %
\newcommand{\BlockOutageMax}{3.54}  %
\newcommand{\BlockTrials}{three}  %
\newcommand{\ReturnMin}{10}  %
\newcommand{\ReturnMax}{13}  %
\newcommand{\ReturnTrials}{six}  %
\newcommand{\FreezeUdp}{3.69}  %
\newcommand{\FreezeTcp}{6.53}  %
\newcommand{\BenchTcpDownMin}{69}  %
\newcommand{\BenchTcpDownMax}{75}  %
\newcommand{\BenchTcpUp}{93}  %
\newcommand{\OffsetHoldsLoss}{10}  %
\newcommand{\OffsetHoldsDelay}{15}  %
\newcommand{\OffsetMovesLoss}{20}  %
\newcommand{\OffsetMovesDelay}{20}  %
\newcommand{\NoiseMovesLoss}{40}  %
\newcommand{\NoOffsetLoss}{1}  %
\newcommand{\NoOffsetDelay}{2}  %
\newcommand{\NoOffsetAfter}{7.9}  %
\newcommand{\RocketMcsMin}{3}  %
\newcommand{\RocketMcsMax}{4}
\newcommand{\RocketOneStreamPct}{83}  %
\newcommand{\BulletMcs}{8}  %
\newcommand{\ChainZeroDbm}{\ensuremath{-}94}  %
\newcommand{\ChainOneDbm}{\ensuremath{-}70}  %
\newcommand{\CapStopRocket}{29.2}  %
\newcommand{\CapStopBullet}{64.7}  %
\newcommand{\EthRocket}{100}  %
\newcommand{\TempRoverFirst}{56}  %
\newcommand{\TempRoverMax}{80}  %
\newcommand{\TempRoverMaxAt}{718}  %
\newcommand{\TempBaseFirst}{55}  %
\newcommand{\TempBaseMax}{61}  %
\newcommand{\DriveStart}{170}  %
\newcommand{\DriveMax}{1{,}339}  %
\newcommand{\DriveMinutes}{28}  %
\newcommand{\RocketLost}{771}  %
\newcommand{\RocketBlip}{958}  %
\newcommand{\BulletLost}{1{,}037}  %
\newcommand{\BulletLostKm}{1.04}  %
\newcommand{\LastPing}{1{,}085}  %
\newcommand{\LastRtcm}{1{,}068}  %
\newcommand{\RtkTo}{1{,}117}  %
\newcommand{\RtkToPlot}{1117}  %
\newcommand{\DenseTo}{775}  %
\newcommand{\PlExpRocket}{2.14}  %

\newcommand{\PlExpBullet}{2.65}

\newcommand{\ThetaMin}{0.1}  %
\newcommand{\ThetaMax}{0.8}  %
\newcommand{\ShortfallMin}{28.5}  %
\newcommand{\ShortfallMax}{37.2}  %
\newcommand{\ShortfallLosMin}{29.4}  %
\newcommand{\ShortfallLosMax}{37.2}  %
\newcommand{\ShortfallObsMin}{28.5}  %
\newcommand{\ShortfallObsMax}{34.8}  %
\newcommand{\PatternLossMin}{4.2}  %
\newcommand{\PatternLossMax}{7.1}  %
\newcommand{\ThetaOutStart}{7.5}  %
\newcommand{\ThetaOutStartAt}{15}  %
\newcommand{\ThetaOutEnd}{1.1}  %
\newcommand{\ThetaOutEndAt}{133}  %
\newcommand{\DipOneFrom}{365}  %
\newcommand{\DipOneTo}{475}  %
\newcommand{\DipOneDepth}{0.7}  %
\newcommand{\DipOneMid}{420}  %
\newcommand{\DipTwoFrom}{625}  %
\newcommand{\DipTwoTo}{735}  %
\newcommand{\DipTwoDepth}{0.8}  %
\newcommand{\DipTwoMid}{680}  %
\newcommand{\DipOneRocket}{\ensuremath{-}83.5}  %
\newcommand{\DipOneBullet}{\ensuremath{-}75.0}  %
\newcommand{\BaseMsl}{5.0}  %
\newcommand{\BaseMslPlot}{5.01}  %
\newcommand{\PingLossIdleMax}{1.9}  %
\newcommand{\SmoothIdleMin}{91.7}  %
\newcommand{\SmoothLoad}{54.3}  %
\newcommand{\RtcmLoad}{80.8}  %
\newcommand{\AgeLoad}{87.9}  %
\newcommand{\PingLossEdge}{53.1}  %
\newcommand{\BaseAvgSeconds}{20}  %
\newcommand{\BaseStdN}{6.7}  %
\newcommand{\BaseStdE}{9.4}  %
\newcommand{\BaseStdU}{1.3}  %
\newcommand{\BaseAbs}{1}  %
\newcommand{\StaticH}{1.4}  %
\newcommand{\StaticV}{1.8}  %
\newcommand{\NearH}{1.7}  %
\newcommand{\NearV}{2.9}  %
\newcommand{\FarH}{1.9}  %
\newcommand{\FarV}{3.1}  %
\newcommand{\PctlH}{2.4}  %
\newcommand{\PctlV}{4.0}  %
\newcommand{\ShareFloat}{81.7}  %
\newcommand{\ShareNone}{17.8}
\newcommand{\ShareFixed}{0.5}
\newcommand{\SpeedMedian}{0.87}  %
\newcommand{\SpeedPctl}{0.96}  %
\newcommand{\SpeedMax}{1.09}  %
\newcommand{\RtcmLatency}{246}  %
\newcommand{\RtcmLatencyPctl}{339}  %
\newcommand{\RtcmRtt}{2.6}  %
\newcommand{\VideoRateStop}{0.26}  %
\newcommand{\VideoRatePctl}{2.8}  %
\newcommand{\BothUpSamples}{914}  %
\newcommand{\OnRocketPct}{99.2}  %
\newcommand{\OnLowerPct}{93.7}  %
\newcommand{\UsedMean}{24.3}  %
\newcommand{\BestMean}{50.9}  %
\newcommand{\MismatchSeconds}{35}  %
\newcommand{\MismatchPct}{2.6}  %
\newcommand{\MismatchWindow}{22}  %
\newcommand{\SwitchesTotal}{24}  %
\newcommand{\EpisodesWord}{three}  %
\newcommand{\EpOneSwitches}{6}  %
\newcommand{\EpOneSpan}{9}  %
\newcommand{\EpOneAt}{105}  %
\newcommand{\EpOneRtcmGap}{3.7}  %
\newcommand{\EpTwoSwitches}{10}  %
\newcommand{\EpTwoSpan}{48}  %
\newcommand{\EpTwoRtcmGap}{19.2}  %
\newcommand{\EpTwoVideo}{57.3}  %
\newcommand{\EpThreeSwitches}{8}  %
\newcommand{\EpThreeSpan}{12}  %
\newcommand{\EpThreeAt}{953}  %
\newcommand{\FieldTcpDown}{10.6}  %
\newcommand{\FieldTcpUp}{5.6}  %
\newcommand{\FieldTcpFrom}{54}  %
\newcommand{\FieldTcpTo}{107}  %
\newcommand{\EdgeFrom}{713}  %
\newcommand{\EdgeTo}{789}  %
\newcommand{\DriveMaxKm}{1.3}  %
\newcommand{\RocketLostBeyondLos}{171}  %
\newcommand{\CapRocketTo}{700}  %
\newcommand{\StopDistMedian}{2.6}  %
\newcommand{\StopDistMax}{3.3}  %
}{}
\IfFileExists{generated/config_macros.tex}{\newcommand{\CmdWindow}{500}  %
\newcommand{\CmdStopTimeout}{3}  %
\newcommand{\CtrlSwitchMean}{541}  %
\newcommand{\CtrlSwitchCILo}{507}  %
\newcommand{\CtrlSwitchCIHi}{575}  %
\newcommand{\CtrlSwitchMin}{482}  %
\newcommand{\CtrlSwitchMax}{618}  %
\newcommand{\CtrlTrials}{ten}  %
\newcommand{\HelloInterval}{0.5}  %
\newcommand{\HelloMissAfter}{0.85}  %
\newcommand{\CostClean}{256}  %
\newcommand{\CostOneLost}{341}  %
\newcommand{\CostTwoLost}{512}  %
\newcommand{\CostThreeLost}{1024}  %
\newcommand{\HellosToInfinity}{twelve}  %
\newcommand{\RttMin}{10}  %
\newcommand{\RttMax}{120}  %
\newcommand{\RttPenaltyMax}{128}  %
\newcommand{\Offset}{256}  %
\newcommand{\OffsetThreshold}{512}  %
\newcommand{\SmoothHalfLife}{2}  %
\newcommand{\FlapLo}{10}  %
\newcommand{\FlapHi}{20}  %
\newcommand{\PhaseLo}{15}  %
\newcommand{\PhaseHi}{30}  %
\newcommand{\EchoInterval}{10}  %
\newcommand{\SetupAt}{11}  %
\newcommand{\StopFrom}{133}  %
\newcommand{\StopTo}{164}  %
\newcommand{\StopMinutes}{13}  %
\newcommand{\LosTo}{600}  %
\newcommand{\SiteLat}{23.849}  %
\newcommand{\SiteLon}{90.506}  %
\newcommand{\PollRadios}{2}  %
\newcommand{\PollPing}{5}  %
\newcommand{\BackoffLo}{2}  %
\newcommand{\BackoffHi}{5}  %
\newcommand{\RoverEpochRate}{5}  %
\newcommand{\LoadRate}{20}  %
\newcommand{\LoadEvery}{3}  %
\newcommand{\LoadFrom}{575}  %
\newcommand{\LoadTo}{935}  %
\newcommand{\LogVideoW}{640}  %
\newcommand{\LogVideoH}{480}  %
\newcommand{\Fps}{30}  %
\newcommand{\SmoothFrames}{25}  %
\newcommand{\AgeLimit}{2}  %
\newcommand{\RtcmNominal}{1{,}000}  %
\newcommand{\BinM}{50}  %
\newcommand{\HeightBinM}{10}  %
\newcommand{\FitFrom}{170}  %
\newcommand{\RocketFreq}{5200}  %
\newcommand{\BulletFreq}{2462}  %
\newcommand{\ChanWidth}{20}  %
\newcommand{\RocketTx}{21}  %
\newcommand{\BulletTxBase}{15}  %
\newcommand{\BulletTxRover}{10}  %
\newcommand{\TelemTx}{30}  %
\newcommand{\SectorGain}{19}  %
\newcommand{\SectorAz}{123}  %
\newcommand{\SectorEl}{4}  %
\newcommand{\SectorLowEdge}{1}  %
\newcommand{\SectorMechTilt}{5}  %
\newcommand{\ElecDowntilt}{2}  %
\newcommand{\SectorBeam}{3}  %
\newcommand{\OmniGain}{13}  %
\newcommand{\OmniEl}{7}  %
\newcommand{\OmniBandLo}{5.45}  %
\newcommand{\OmniBandHi}{5.85}  %
\newcommand{\BulletAntGain}{8}  %
\newcommand{\PNineGain}{6}  %
\newcommand{\SatelGain}{2}  %
\newcommand{\MastHeight}{3}  %
\newcommand{\RoverRocketHeight}{1.5}  %
\newcommand{\RoverBulletHeight}{2.0}  %
\newcommand{\RoverPNineHeight}{3.0}  %
\newcommand{\RoverSatelHeight}{2.0}  %
\newcommand{\MastHeightOne}{3.0}  %
\newcommand{\PNineLo}{902}  %
\newcommand{\PNineHi}{928}  %
\newcommand{\SatelChan}{25}  %
\newcommand{\RocketGHz}{5.2}  %
\newcommand{\BandTwoFour}{2.4}  %
\newcommand{\BandFive}{5}  %
\newcommand{\BandNine}{900}  %
\newcommand{\BandFourThree}{433}  %
\newcommand{\CiLevel}{95}  %
\newcommand{\PctlLevel}{95}  %
\newcommand{\FitRef}{100}  %
\newcommand{\FsMiscLoss}{1}  %
\newcommand{\PatternK}{12}  %
\newcommand{\PatternFloor}{20}  %
\newcommand{\HelloHistory}{sixteen}  %
\newcommand{\LowerMargin}{10}  %
\newcommand{\MovingSpeed}{0.2}  %
\newcommand{\NearFrom}{150}  %
\newcommand{\FarTo}{1{,}100}  %
\newcommand{\AntVexag}{five}  %
\newcommand{\AntSketchEx}{three}  %
\newcommand{\CompRangeLo}{600}  %
\newcommand{\CompRangeHi}{900}  %
\newcommand{\UrcRangeKm}{1}  %
\newcommand{\SniderPingGap}{3}  %
\newcommand{\ByuRate}{2.53}  %
\newcommand{\ByuRange}{1}  %
\newcommand{\WvuBand}{70}  %
\newcommand{\GuelSingleLo}{68}  %
\newcommand{\GuelSingleHi}{97}  %
\newcommand{\GuelMulti}{99.8}  %
\newcommand{\LauJitter}{1}  %
\newcommand{\AbolChanges}{7.8}  %
\newcommand{\NeumannNodes}{50}  %
\newcommand{\ArduRc}{1}  %
\newcommand{\ArduGcs}{5}  %
\newcommand{\PxLoss}{10}  %
}{}
\IfFileExists{generated/table_ablation.tex}{\newcommand{\TabAblation}{%
\multicolumn{4}{@{}l}{\emph{Without the +256 offset}} \\
1\% loss, +2\,ms & 0.7\% & -- & Bullet after 7.9\,s \\
15\% bursty loss & 1.3\% & -- & Stays on Rocket \\
30\% bursty loss & 2.6\% & -- & Bullet after 2.8\,s \\
\midrule
\multicolumn{4}{@{}l}{\emph{With the +256 offset}} \\
1\% loss, +2\,ms & 1.5\% & 11.2 & Stays on Rocket \\
3\% loss, +5\,ms & 5.3\% & 17.7 & Stays on Rocket \\
5\% loss, +10\,ms & 8.8\% & 25.9 & Stays on Rocket \\
10\% loss, +15\,ms & 18.7\% & 37.8 & Stays on Rocket \\
20\% loss, +20\,ms & 35.5\% & 45.2 & Bullet after 14.1\,s \\
10\% loss, no delay & 18.4\% & 3.7 & Stays on Rocket \\
20\% loss, no delay & 36.2\% & 3.9 & Stays on Rocket \\
40\% loss, no delay & 29.8\% & 3.7 & Bullet after 6.5\,s, flaps first \\
100\% loss (blocked) & -- & -- & Outage 3.19--3.54\,s, $n=3$ \\
}
}{}
\IfFileExists{generated/table_pathloss.tex}{\newcommand{\TabPathloss}{%
Rocket, rover & 2.14 (1.20--3.09) & \ensuremath{-}66.2 & 2.8 & 0.72 & 12 \\
Bullet, rover & 2.65 (2.04--3.26) & \ensuremath{-}58.0 & 2.0 & 0.89 & 13 \\
Rocket, base, chain 1 & 2.32 (1.33--3.31) & \ensuremath{-}62.8 & 3.2 & 0.71 & 13 \\
Bullet, base & 2.38 (1.77--3.00) & \ensuremath{-}63.7 & 2.0 & 0.87 & 13 \\
}
}{}
\IfFileExists{generated/table_service.tex}{\newcommand{\TabService}{%
133--164 (stop) & None & 0.0\% & 2.1 & 99.5\% & 100.0\% & 100.0\% \\
150--350 & None & 0.0\% & 4.5 & 95.7\% & 100.0\% & 100.0\% \\
350--575 & None & 1.9\% & 10.0 & 93.7\% & 100.0\% & 91.7\% \\
575--935 & 20\,Mb/s & 3.0\% & 12.7 & 80.8\% & 87.9\% & 54.3\% \\
935--1{,}000 & None & 5.0\% & 4.2 & 84.4\% & 92.1\% & 78.6\% \\
1{,}000--1{,}100 & None & 53.1\% & 26.4 & 23.5\% & 27.6\% & 26.4\% \\
$>$1{,}100 & None & 100.0\% & -- & 0.0\% & 0.0\% & 0.0\% \\
}
}{}
\providecommand{\TabAblation}{\tbd{run analysis/paper\_macros.py}\\}
\providecommand{\TabPathloss}{\tbd{run analysis/paper\_macros.py}\\}
\providecommand{\TabService}{\tbd{run analysis/paper\_macros.py}\\}
\providecommand{\AbolChanges}{\tbd{AbolChanges}}
\providecommand{\AgeLimit}{\tbd{AgeLimit}}
\providecommand{\AgeLoad}{\tbd{AgeLoad}}
\providecommand{\AntSketchEx}{\tbd{AntSketchEx}}
\providecommand{\AntVexag}{\tbd{AntVexag}}
\providecommand{\ArduGcs}{\tbd{ArduGcs}}
\providecommand{\ArduRc}{\tbd{ArduRc}}
\providecommand{\BackoffHi}{\tbd{BackoffHi}}
\providecommand{\BackoffLo}{\tbd{BackoffLo}}
\providecommand{\BandFive}{\tbd{BandFive}}
\providecommand{\BandFourThree}{\tbd{BandFourThree}}
\providecommand{\BandNine}{\tbd{BandNine}}
\providecommand{\BandTwoFour}{\tbd{BandTwoFour}}
\providecommand{\BaseAbs}{\tbd{BaseAbs}}
\providecommand{\BaseAvgSeconds}{\tbd{BaseAvgSeconds}}
\providecommand{\BaseMsl}{\tbd{BaseMsl}}
\providecommand{\BaseMslPlot}{\tbd{BaseMslPlot}}
\providecommand{\BaseStdE}{\tbd{BaseStdE}}
\providecommand{\BaseStdN}{\tbd{BaseStdN}}
\providecommand{\BaseStdU}{\tbd{BaseStdU}}
\providecommand{\BenchTcpDownMax}{\tbd{BenchTcpDownMax}}
\providecommand{\BenchTcpDownMin}{\tbd{BenchTcpDownMin}}
\providecommand{\BenchTcpUp}{\tbd{BenchTcpUp}}
\providecommand{\BestMean}{\tbd{BestMean}}
\providecommand{\BinM}{\tbd{BinM}}
\providecommand{\BlockOutageMax}{\tbd{BlockOutageMax}}
\providecommand{\BlockOutageMin}{\tbd{BlockOutageMin}}
\providecommand{\BlockTrials}{\tbd{BlockTrials}}
\providecommand{\BothUpSamples}{\tbd{BothUpSamples}}
\providecommand{\BulletAntGain}{\tbd{BulletAntGain}}
\providecommand{\BulletFreq}{\tbd{BulletFreq}}
\providecommand{\BulletLost}{\tbd{BulletLost}}
\providecommand{\BulletLostKm}{\tbd{BulletLostKm}}
\providecommand{\BulletMcs}{\tbd{BulletMcs}}
\providecommand{\BulletTxBase}{\tbd{BulletTxBase}}
\providecommand{\BulletTxRover}{\tbd{BulletTxRover}}
\providecommand{\ByuRange}{\tbd{ByuRange}}
\providecommand{\ByuRate}{\tbd{ByuRate}}
\providecommand{\CapRocketTo}{\tbd{CapRocketTo}}
\providecommand{\CapStopBullet}{\tbd{CapStopBullet}}
\providecommand{\CapStopRocket}{\tbd{CapStopRocket}}
\providecommand{\ChainOneDbm}{\tbd{ChainOneDbm}}
\providecommand{\ChainZeroDbm}{\tbd{ChainZeroDbm}}
\providecommand{\ChanWidth}{\tbd{ChanWidth}}
\providecommand{\CiLevel}{\tbd{CiLevel}}
\providecommand{\CmdStopTimeout}{\tbd{CmdStopTimeout}}

\providecommand{\CmdWindow}{\tbd{CmdWindow}}
\providecommand{\CompRangeHi}{\tbd{CompRangeHi}}
\providecommand{\CompRangeLo}{\tbd{CompRangeLo}}
\providecommand{\CostClean}{\tbd{CostClean}}
\providecommand{\CostOneLost}{\tbd{CostOneLost}}
\providecommand{\CostThreeLost}{\tbd{CostThreeLost}}
\providecommand{\CostTwoLost}{\tbd{CostTwoLost}}
\providecommand{\CtrlSwitchCIHi}{\tbd{CtrlSwitchCIHi}}
\providecommand{\CtrlSwitchCILo}{\tbd{CtrlSwitchCILo}}
\providecommand{\CtrlSwitchMax}{\tbd{CtrlSwitchMax}}
\providecommand{\CtrlSwitchMean}{\tbd{CtrlSwitchMean}}
\providecommand{\CtrlSwitchMin}{\tbd{CtrlSwitchMin}}
\providecommand{\CtrlTrials}{\tbd{CtrlTrials}}
\providecommand{\DenseTo}{\tbd{DenseTo}}
\providecommand{\DipOneBullet}{\tbd{DipOneBullet}}
\providecommand{\DipOneDepth}{\tbd{DipOneDepth}}
\providecommand{\DipOneFrom}{\tbd{DipOneFrom}}
\providecommand{\DipOneMid}{\tbd{DipOneMid}}
\providecommand{\DipOneRocket}{\tbd{DipOneRocket}}
\providecommand{\DipOneTo}{\tbd{DipOneTo}}
\providecommand{\DipTwoDepth}{\tbd{DipTwoDepth}}
\providecommand{\DipTwoFrom}{\tbd{DipTwoFrom}}
\providecommand{\DipTwoMid}{\tbd{DipTwoMid}}
\providecommand{\DipTwoTo}{\tbd{DipTwoTo}}
\providecommand{\DriveMax}{\tbd{DriveMax}}
\providecommand{\DriveMaxKm}{\tbd{DriveMaxKm}}
\providecommand{\DriveMinutes}{\tbd{DriveMinutes}}
\providecommand{\DriveStart}{\tbd{DriveStart}}
\providecommand{\EchoInterval}{\tbd{EchoInterval}}
\providecommand{\EdgeFrom}{\tbd{EdgeFrom}}
\providecommand{\EdgeTo}{\tbd{EdgeTo}}
\providecommand{\ElecDowntilt}{\tbd{ElecDowntilt}}
\providecommand{\EpOneAt}{\tbd{EpOneAt}}
\providecommand{\EpOneRtcmGap}{\tbd{EpOneRtcmGap}}
\providecommand{\EpOneSpan}{\tbd{EpOneSpan}}
\providecommand{\EpOneSwitches}{\tbd{EpOneSwitches}}

\providecommand{\EpThreeAt}{\tbd{EpThreeAt}}

\providecommand{\EpThreeSpan}{\tbd{EpThreeSpan}}
\providecommand{\EpThreeSwitches}{\tbd{EpThreeSwitches}}

\providecommand{\EpTwoRtcmGap}{\tbd{EpTwoRtcmGap}}
\providecommand{\EpTwoSpan}{\tbd{EpTwoSpan}}
\providecommand{\EpTwoSwitches}{\tbd{EpTwoSwitches}}
\providecommand{\EpTwoVideo}{\tbd{EpTwoVideo}}
\providecommand{\EpisodesWord}{\tbd{EpisodesWord}}
\providecommand{\EthRocket}{\tbd{EthRocket}}
\providecommand{\FarH}{\tbd{FarH}}
\providecommand{\FarTo}{\tbd{FarTo}}
\providecommand{\FarV}{\tbd{FarV}}
\providecommand{\FieldTcpDown}{\tbd{FieldTcpDown}}
\providecommand{\FieldTcpFrom}{\tbd{FieldTcpFrom}}
\providecommand{\FieldTcpTo}{\tbd{FieldTcpTo}}
\providecommand{\FieldTcpUp}{\tbd{FieldTcpUp}}
\providecommand{\FitFrom}{\tbd{FitFrom}}
\providecommand{\FitRef}{\tbd{FitRef}}
\providecommand{\FlapHi}{\tbd{FlapHi}}
\providecommand{\FlapLo}{\tbd{FlapLo}}
\providecommand{\Fps}{\tbd{Fps}}
\providecommand{\FreezeTcp}{\tbd{FreezeTcp}}
\providecommand{\FreezeUdp}{\tbd{FreezeUdp}}
\providecommand{\FsMiscLoss}{\tbd{FsMiscLoss}}
\providecommand{\GuelMulti}{\tbd{GuelMulti}}
\providecommand{\GuelSingleHi}{\tbd{GuelSingleHi}}
\providecommand{\GuelSingleLo}{\tbd{GuelSingleLo}}
\providecommand{\HeightBinM}{\tbd{HeightBinM}}

\providecommand{\HelloHistory}{\tbd{HelloHistory}}
\providecommand{\HelloInterval}{\tbd{HelloInterval}}
\providecommand{\HelloMissAfter}{\tbd{HelloMissAfter}}
\providecommand{\HellosToInfinity}{\tbd{HellosToInfinity}}
\providecommand{\LastPing}{\tbd{LastPing}}
\providecommand{\LastRtcm}{\tbd{LastRtcm}}
\providecommand{\LauJitter}{\tbd{LauJitter}}
\providecommand{\LoadEvery}{\tbd{LoadEvery}}
\providecommand{\LoadFrom}{\tbd{LoadFrom}}
\providecommand{\LoadRate}{\tbd{LoadRate}}
\providecommand{\LoadTo}{\tbd{LoadTo}}
\providecommand{\LogVideoH}{\tbd{LogVideoH}}
\providecommand{\LogVideoW}{\tbd{LogVideoW}}
\providecommand{\LosTo}{\tbd{LosTo}}
\providecommand{\LowerMargin}{\tbd{LowerMargin}}
\providecommand{\MastHeight}{\tbd{MastHeight}}
\providecommand{\MastHeightOne}{\tbd{MastHeightOne}}
\providecommand{\MismatchPct}{\tbd{MismatchPct}}
\providecommand{\MismatchSeconds}{\tbd{MismatchSeconds}}
\providecommand{\MismatchWindow}{\tbd{MismatchWindow}}
\providecommand{\MovingSpeed}{\tbd{MovingSpeed}}
\providecommand{\NearFrom}{\tbd{NearFrom}}
\providecommand{\NearH}{\tbd{NearH}}
\providecommand{\NearV}{\tbd{NearV}}
\providecommand{\NeumannNodes}{\tbd{NeumannNodes}}
\providecommand{\NoOffsetAfter}{\tbd{NoOffsetAfter}}
\providecommand{\NoOffsetDelay}{\tbd{NoOffsetDelay}}
\providecommand{\NoOffsetLoss}{\tbd{NoOffsetLoss}}

\providecommand{\NoiseMovesLoss}{\tbd{NoiseMovesLoss}}
\providecommand{\Offset}{\tbd{Offset}}
\providecommand{\OffsetHoldsDelay}{\tbd{OffsetHoldsDelay}}
\providecommand{\OffsetHoldsLoss}{\tbd{OffsetHoldsLoss}}

\providecommand{\OffsetMovesDelay}{\tbd{OffsetMovesDelay}}
\providecommand{\OffsetMovesLoss}{\tbd{OffsetMovesLoss}}
\providecommand{\OffsetThreshold}{\tbd{OffsetThreshold}}
\providecommand{\OmniBandHi}{\tbd{OmniBandHi}}
\providecommand{\OmniBandLo}{\tbd{OmniBandLo}}
\providecommand{\OmniEl}{\tbd{OmniEl}}
\providecommand{\OmniGain}{\tbd{OmniGain}}
\providecommand{\OnLowerPct}{\tbd{OnLowerPct}}
\providecommand{\OnRocketPct}{\tbd{OnRocketPct}}
\providecommand{\PNineGain}{\tbd{PNineGain}}
\providecommand{\PNineHi}{\tbd{PNineHi}}
\providecommand{\PNineLo}{\tbd{PNineLo}}
\providecommand{\PatternFloor}{\tbd{PatternFloor}}
\providecommand{\PatternK}{\tbd{PatternK}}
\providecommand{\PatternLossMax}{\tbd{PatternLossMax}}
\providecommand{\PatternLossMin}{\tbd{PatternLossMin}}
\providecommand{\PctlH}{\tbd{PctlH}}
\providecommand{\PctlLevel}{\tbd{PctlLevel}}
\providecommand{\PctlV}{\tbd{PctlV}}
\providecommand{\PhaseHi}{\tbd{PhaseHi}}
\providecommand{\PhaseLo}{\tbd{PhaseLo}}
\providecommand{\PingLossEdge}{\tbd{PingLossEdge}}
\providecommand{\PingLossIdleMax}{\tbd{PingLossIdleMax}}
\providecommand{\PlExpBullet}{\tbd{PlExpBullet}}

\providecommand{\PlExpRocket}{\tbd{PlExpRocket}}

\providecommand{\PollPing}{\tbd{PollPing}}
\providecommand{\PollRadios}{\tbd{PollRadios}}

\providecommand{\PowerOffOutage}{\tbd{PowerOffOutage}}
\providecommand{\PxLoss}{\tbd{PxLoss}}

\providecommand{\ReturnMax}{\tbd{ReturnMax}}
\providecommand{\ReturnMin}{\tbd{ReturnMin}}
\providecommand{\ReturnTrials}{\tbd{ReturnTrials}}
\providecommand{\RocketBlip}{\tbd{RocketBlip}}
\providecommand{\RocketFreq}{\tbd{RocketFreq}}
\providecommand{\RocketGHz}{\tbd{RocketGHz}}
\providecommand{\RocketLost}{\tbd{RocketLost}}
\providecommand{\RocketLostBeyondLos}{\tbd{RocketLostBeyondLos}}
\providecommand{\RocketMcsMax}{\tbd{RocketMcsMax}}
\providecommand{\RocketMcsMin}{\tbd{RocketMcsMin}}
\providecommand{\RocketOneStreamPct}{\tbd{RocketOneStreamPct}}
\providecommand{\RocketTx}{\tbd{RocketTx}}

\providecommand{\RoverBulletHeight}{\tbd{RoverBulletHeight}}
\providecommand{\RoverEpochRate}{\tbd{RoverEpochRate}}
\providecommand{\RoverPNineHeight}{\tbd{RoverPNineHeight}}
\providecommand{\RoverRocketHeight}{\tbd{RoverRocketHeight}}
\providecommand{\RoverSatelHeight}{\tbd{RoverSatelHeight}}
\providecommand{\RtcmLatency}{\tbd{RtcmLatency}}
\providecommand{\RtcmLatencyPctl}{\tbd{RtcmLatencyPctl}}
\providecommand{\RtcmLoad}{\tbd{RtcmLoad}}
\providecommand{\RtcmNominal}{\tbd{RtcmNominal}}

\providecommand{\RtcmRtt}{\tbd{RtcmRtt}}
\providecommand{\RtkTo}{\tbd{RtkTo}}
\providecommand{\RtkToPlot}{\tbd{RtkToPlot}}
\providecommand{\RttMax}{\tbd{RttMax}}
\providecommand{\RttMin}{\tbd{RttMin}}
\providecommand{\RttPenaltyMax}{\tbd{RttPenaltyMax}}
\providecommand{\SatelChan}{\tbd{SatelChan}}
\providecommand{\SatelGain}{\tbd{SatelGain}}
\providecommand{\SectorAz}{\tbd{SectorAz}}
\providecommand{\SectorBeam}{\tbd{SectorBeam}}
\providecommand{\SectorEl}{\tbd{SectorEl}}
\providecommand{\SectorGain}{\tbd{SectorGain}}
\providecommand{\SectorLowEdge}{\tbd{SectorLowEdge}}
\providecommand{\SectorMechTilt}{\tbd{SectorMechTilt}}
\providecommand{\SetupAt}{\tbd{SetupAt}}
\providecommand{\ShareFixed}{\tbd{ShareFixed}}
\providecommand{\ShareFloat}{\tbd{ShareFloat}}
\providecommand{\ShareNone}{\tbd{ShareNone}}
\providecommand{\ShortfallLosMax}{\tbd{ShortfallLosMax}}
\providecommand{\ShortfallLosMin}{\tbd{ShortfallLosMin}}
\providecommand{\ShortfallMax}{\tbd{ShortfallMax}}
\providecommand{\ShortfallMin}{\tbd{ShortfallMin}}
\providecommand{\ShortfallObsMax}{\tbd{ShortfallObsMax}}
\providecommand{\ShortfallObsMin}{\tbd{ShortfallObsMin}}

\providecommand{\SiteLat}{\tbd{SiteLat}}
\providecommand{\SiteLon}{\tbd{SiteLon}}
\providecommand{\SmoothFrames}{\tbd{SmoothFrames}}
\providecommand{\SmoothHalfLife}{\tbd{SmoothHalfLife}}
\providecommand{\SmoothIdleMin}{\tbd{SmoothIdleMin}}
\providecommand{\SmoothLoad}{\tbd{SmoothLoad}}
\providecommand{\SniderPingGap}{\tbd{SniderPingGap}}
\providecommand{\SpeedMax}{\tbd{SpeedMax}}
\providecommand{\SpeedMedian}{\tbd{SpeedMedian}}
\providecommand{\SpeedPctl}{\tbd{SpeedPctl}}
\providecommand{\StaticH}{\tbd{StaticH}}
\providecommand{\StaticV}{\tbd{StaticV}}
\providecommand{\StopDistMax}{\tbd{StopDistMax}}
\providecommand{\StopDistMedian}{\tbd{StopDistMedian}}
\providecommand{\StopFrom}{\tbd{StopFrom}}
\providecommand{\StopMinutes}{\tbd{StopMinutes}}
\providecommand{\StopTo}{\tbd{StopTo}}
\providecommand{\SwitchesTotal}{\tbd{SwitchesTotal}}
\providecommand{\TelemTx}{\tbd{TelemTx}}
\providecommand{\TempBaseFirst}{\tbd{TempBaseFirst}}
\providecommand{\TempBaseMax}{\tbd{TempBaseMax}}
\providecommand{\TempRoverFirst}{\tbd{TempRoverFirst}}
\providecommand{\TempRoverMax}{\tbd{TempRoverMax}}
\providecommand{\TempRoverMaxAt}{\tbd{TempRoverMaxAt}}
\providecommand{\ThetaMax}{\tbd{ThetaMax}}
\providecommand{\ThetaMin}{\tbd{ThetaMin}}
\providecommand{\ThetaOutEnd}{\tbd{ThetaOutEnd}}
\providecommand{\ThetaOutEndAt}{\tbd{ThetaOutEndAt}}
\providecommand{\ThetaOutStart}{\tbd{ThetaOutStart}}
\providecommand{\ThetaOutStartAt}{\tbd{ThetaOutStartAt}}
\providecommand{\UrcRangeKm}{\tbd{UrcRangeKm}}
\providecommand{\UsedMean}{\tbd{UsedMean}}
\providecommand{\VideoRatePctl}{\tbd{VideoRatePctl}}
\providecommand{\VideoRateStop}{\tbd{VideoRateStop}}
\providecommand{\WvuBand}{\tbd{WvuBand}}

\definecolor{rocketc}{HTML}{2A78D6}
\definecolor{bulletc}{HTML}{EB6834}
\definecolor{archbg}{HTML}{F4F5F7}   %

\begin{document}

\title{Band-Separated Redundant Communication for a Mars-Analog Rover}

\author{%
\IEEEauthorblockN{Zamiul Rashid, Akid Anis, Tanzim Ahmed, Md.\ Golam Rabiul Alam$^{*}$, Md.\ Khalilur Rhaman$^{*}$\thanks{$^{*}$Supervisors.}}
\IEEEauthorblockA{BRAC University, Dhaka, Bangladesh\\
\texttt{\{zamiul.rashid, akid.anis, tanzim.ahmed3\}@g.bracu.ac.bd}\\
\texttt{\{rabiul.alam, khalilur\}@bracu.ac.bd}}}

\maketitle

\begin{abstract}
Teams in the University Rover Challenge (URC) drive their rovers remotely and often out of sight. The organizers name communication failure as the most common reason a team falls short. A rover therefore needs bandwidth for several camera views and a command path that survives the loss of one radio. We describe and field test the communication stack of the BRACU Mongol-Tori rover, which keeps vision and control on separate radios and bands. Video and RTK corrections travel over a \RocketGHz\,GHz bridge with a \BandTwoFour\,GHz backup, and the Babel routing protocol chooses between them. Every command goes out on two sub-gigahertz radios at once, and the rover uses the backup copy only when the primary falls silent. We tested the backhaul on the bench with a power cut and injected packet loss. We also drove the rover more than a kilometer from the base, and line of sight ended about halfway. On the bench, traffic reached the backup within a few seconds. A metric offset kept it on a primary that lost one packet in ten. In the field, the primary failed at \RocketLost\,m and the backup carried video and corrections to just over a kilometer. The drive also exposed a limit of Babel, which judges a link only by lost Hello messages. It kept traffic on a primary with a faulty receiver, although the backup reported twice the capacity, and it flapped at the edge of range. We close with lessons on antenna aiming, link metrics and logging.
\end{abstract}


\section{Introduction}
\label{sec:intro}
The University Rover Challenge (URC) asks student teams to build Mars analog rovers and to operate them at the Mars Desert Research Station in Utah. Operators drive from command stations that block their view of the rover~\cite{urc_requirements}. Rovers may travel up to \UrcRangeKm\,km, and in the Delivery Mission line of sight is not guaranteed for more than half the course. One task sits in a remote canyon where communication problems are expected~\cite{urc_delivery_2025}. The organizers call communication failures the most common problem that keeps a team from contending~\cite{urc_advice}.

A rover built for these events therefore has two communication needs that pull in different directions. Operators need several live camera views and a stream of RTK corrections for precise positioning. These need an IP link of several megabits per second, which in practice means a Wi-Fi radio. Such radios lose range quickly once trees or terrain block the line of sight~\cite{snider2018urc,byu_itc2018}. Commands need little bandwidth but must arrive within milliseconds. Low-rate sub-gigahertz radios suit them and reach farther. Carrying both on one link couples their failures, because a saturated or fading video link also delays the commands.

Published team reports describe a single link, or a backup link whose transition is not measured (Section~\ref{sec:related}). Routing protocols such as Babel can choose between two links automatically. They have been compared as multi-hop mesh protocols, but not as the switch between the primary and the backup link of one vehicle. A team that chooses a design thus has little measured evidence on how fast a backup takes over, or on what makes it take over.

The BRACU Mongol-Tori stack keeps vision and control on separate radios in separate bands. Video and RTK corrections travel over a \RocketGHz\,GHz bridge, the Rocket link. A \BandTwoFour\,GHz bridge, the Bullet link, is its backup. Babel runs on a router at each end and chooses between the two links. Every command goes out on a \BandNine\,MHz and a \BandFourThree\,MHz radio at once. The rover's controller acts on the \BandNine\,MHz copy unless none has arrived for about \CmdWindow\,ms. Each of these mechanisms is established practice. Our contribution is a measured account of how they behave together on a Mars-analog rover, both when a link fails and when it only degrades.

We measured the stack in three campaigns. All ran in Dhaka, Bangladesh, and none at a competition site. The command switchover was timed in \CtrlTrials{} trials in which the \BandNine\,MHz radio was powered off abruptly. On the bench we powered off the primary video radio. In separate runs, Linux netem injected rising loss and delay on its router ports while the radios stayed associated. Echoes every \EchoInterval\,ms between base and rover timed each outage, and each router's route was read every second. Runs were made with and without a metric offset that favors the primary. Finally, the rover drove \DriveMaxKm\,km from the base, in line of sight to about \LosTo\,m and behind trees beyond. Along the drive we logged signal and capacity from all four radios and the route in use. We also logged echoes, every RTK epoch and correction frame, and one live video stream. The received signal was then compared with the antenna geometry, using the rover's RTK height and the antenna datasheets (Section~\ref{sec:method}).

Our contribution is fourfold. First, we describe a band separated rover stack in which video and corrections share no radio with control (Section~\ref{sec:system}). We also name the single points of failure that remain. Second, we explain Babel as a selector between a primary and a backup link, checked against the babeld source (Section~\ref{sec:babel}). The account covers how lost Hellos set the link cost and how an offset makes one link primary. It also shows why one route per destination gives failover rather than load balancing. Third, we measure the stack on the bench and in the field (Section~\ref{sec:results}). The bench gives outage times after a power loss and a blockage, and the loss at which the offset gives up the primary. The drive gives range, echo loss, video continuity, correction delivery, RTK accuracy and route choice over \DriveMaxKm\,km. Fourth, we distill lessons for other teams on antenna aiming, link metrics, load and logging (Section~\ref{sec:discussion}).

\section{Related Work}
\label{sec:related}
Snider et al.\ survey eight URC teams, most of which rely on a \BandTwoFour\,GHz link from a tall mast~\cite{snider2018urc}. One surveyed team backs its \BandTwoFour\,GHz video link with a \BandFourThree\,MHz LoRa link for telemetry. Another declares the link lost after two failed pings sent \SniderPingGap\,s apart. Neither reports how long the switch takes. Phoenix carries video on \BandFive\,GHz and commands on \BandNine\,MHz~\cite{zaman2022phoenix}. A multiplexer moves its commands to \BandTwoFour\,GHz on an error or delay in the \BandNine\,MHz band. The paper reports command response times but not how a fault is detected or how long the switch takes. BYU found its \BandTwoFour\,GHz link unreliable out of line of sight~\cite{byu_itc2018}. The team replaced it with one \BandNine\,MHz link and a GPS-steered Yagi antenna and measured \ByuRate\,Mb/s at \ByuRange\,km in line of sight. West Virginia University paired \BandTwoFour\,GHz with \BandNine\,MHz~\cite{wvu_sar2024}. Its later rover added \WvuBand\,cm packet radios for low-rate control and described a seamless transition out of line of sight, without its mechanism or timing~\cite{wvu_sar2025}. Our team's earlier rover sent commands over a \BandTwoFour\,GHz bridge and video on a separate analog transmitter, with no backup for either~\cite{mongoltori2021}. None of these reports measures a switchover.

Güldenring et al.\ flew a search and rescue aircraft with three LTE links~\cite{gueldenring2020multilink}. Single links were available \GuelSingleLo--\GuelSingleHi\% of the time and the combination up to \GuelMulti\%. The combined figure counts any link that answered an echo. It therefore bounds what switching could achieve rather than measuring a switch. Lau and Ang switch a delivery drone between cellular, SMS and satellite links when a heartbeat times out~\cite{lau2021link}. In lab tests their adaptive timeout failed to converge once delay jitter exceeded \LauJitter\,s.

PRP and IEEE 802.1CB avoid switching altogether~\cite{kirrmann2007prp,iec62439_3,ieee8021cb}. Each frame travels over two paths and the receiver discards the duplicate, so recovery is bumpless. Our command radios also duplicate every message. The controller, however, prefers the primary's copy instead of keeping the first to arrive.

Babel~\cite{rfc6126,rfc8966} is a distance-vector protocol for networks with lossy wireless links. Neighbors exchange periodic Hellos, and the specification suggests ETX for the cost of a wireless link. ETX is the inverse product of the forward and reverse delivery ratios of broadcast probes~\cite{decouto2003etx}. A delay extension adds a penalty that grows with round-trip time~\cite{jonglez2014delay,rfc9616}. To avoid oscillation, a router switches routes only when the candidate's metric is no higher and its smoothed metric is lower. No input to the cost reflects signal level or capacity.

Testbed studies compare Babel with OLSR and B.A.T.M.A.N.\ as a mesh protocol. On eight indoor nodes it repaired routes fastest but changed routes up to \AbolChanges{} times a minute~\cite{abolhasan2009realworld}. It had the lowest overhead on small topologies~\cite{murray2010experimental}. On \NeumannNodes{} nodes it healed more slowly than BMX6, partly because of its longer default Hello interval~\cite{neumann2015evaluation}. Its throughput fell steeply as a Raspberry~Pi mesh grew~\cite{turlykozhayeva2024raspberry}. On quadcopters it was the most stable of three protocols, but the switching time was not measured~\cite{guillenperez2021fanet}. We instead use Babel to choose between two parallel links of one vehicle, and we time the switch.

ArduPilot Rover's failsafes act after \ArduRc\,s without RC input or \ArduGcs\,s without a ground station heartbeat~\cite{ardupilot_rover_failsafe}. PX4 waits \PxLoss\,s by default before it declares a data link lost~\cite{px4_safety}. Both act on total loss. Neither moves traffic to another link, and both wait at least twice as long as our window of about \CmdWindow\,ms.

\section{System}
\label{sec:system}
\begin{figure*}[t]
\centering
\setlength{\fboxsep}{7pt}%
\colorbox{archbg}{\includegraphics[width=\dimexpr0.86\textwidth-14pt\relax]{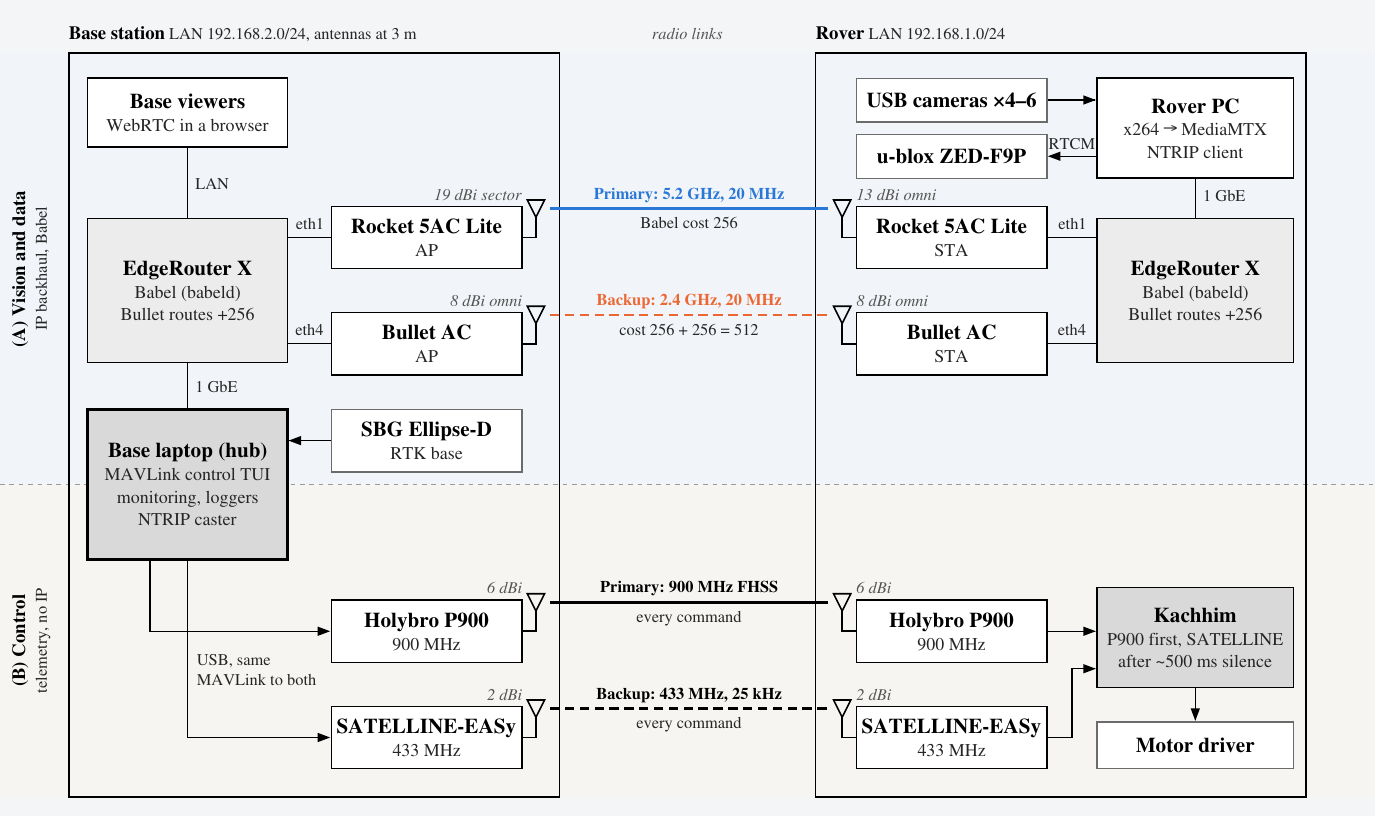}}
\caption{\textbf{Communication architecture:} (A)~Vision and data. Camera streams and RTK corrections travel over the \RocketGHz\,GHz Rocket link, and Babel moves them to the \BandTwoFour\,GHz Bullet link when needed. (B)~Control. The same MAVLink stream goes out on the \BandNine\,MHz P900 and the \BandFourThree\,MHz SATELLINE-EASy. Solid radio links are primary and dashed links are backups. Each antenna symbol carries its gain.}
\label{fig:system_arch}
\end{figure*}

Fig.~\ref{fig:system_arch} shows the stack. The base laptop is its hub. It sends the commands and monitors every subsystem. Video returns onto the base LAN, where any base computer can view it. Table~\ref{tab:links} lists the four radio links. Every base antenna sits at \MastHeight\,m, the URC maximum~\cite{urc_requirements}.

\begin{table}[t]
\centering
\caption{Radio Links as Configured for the Field Drive}
\label{tab:links}
\footnotesize
\setlength{\tabcolsep}{3pt}
\begin{tabularx}{\columnwidth}{@{}L{1.6cm}L{1.6cm}cYc@{}}
\toprule
\textbf{Link} & \textbf{Band} & \textbf{TX} & \textbf{Antenna, base / rover} & \textbf{Height} \\
\midrule
Rocket 5AC Lite & \RocketFreq\,MHz, \ChanWidth\,MHz & \RocketTx{} / \RocketTx & \SectorGain\,dBi sector, \SectorAz$^\circ\times$\SectorEl$^\circ$, beam \SectorBeam$^\circ$ up / \OmniGain\,dBi omni, 360$^\circ\times$\OmniEl$^\circ$ & \MastHeightOne{} / \RoverRocketHeight \\
Bullet AC & \BulletFreq\,MHz, \ChanWidth\,MHz & \BulletTxBase{} / \BulletTxRover & \BulletAntGain\,dBi omni / \BulletAntGain\,dBi omni & \MastHeightOne{} / \RoverBulletHeight \\
P900 & \PNineLo--\PNineHi\,MHz FHSS, one sub-band & \TelemTx & \PNineGain\,dBi & \MastHeightOne{} / \RoverPNineHeight \\
SATELLINE-EASy & \BandFourThree\,MHz, \SatelChan\,kHz 4FSK & \TelemTx & \SatelGain\,dBi & \MastHeightOne{} / \RoverSatelHeight \\
\bottomrule
\end{tabularx}

\vspace{2pt}
{\footnotesize TX power in dBm and height in m, each as base / rover. Beamwidths are horizontal $\times$ vertical, from the datasheets~\cite{ubiquiti_sector,ubiquiti_amo5g13,ubiquiti_rocket,ubiquiti_bullet,px4_p900,satel_easy} (patterns in Fig.~\ref{fig:datasheet_patterns}). The sector's \SectorMechTilt$^\circ$ uptilt and \ElecDowntilt$^\circ$ electrical downtilt aim its beam \SectorBeam$^\circ$ up.}
\end{table}

\begin{figure*}[t]
\centering
\begin{minipage}[t]{0.48\textwidth}
\centering
\includegraphics[height=2.85in]{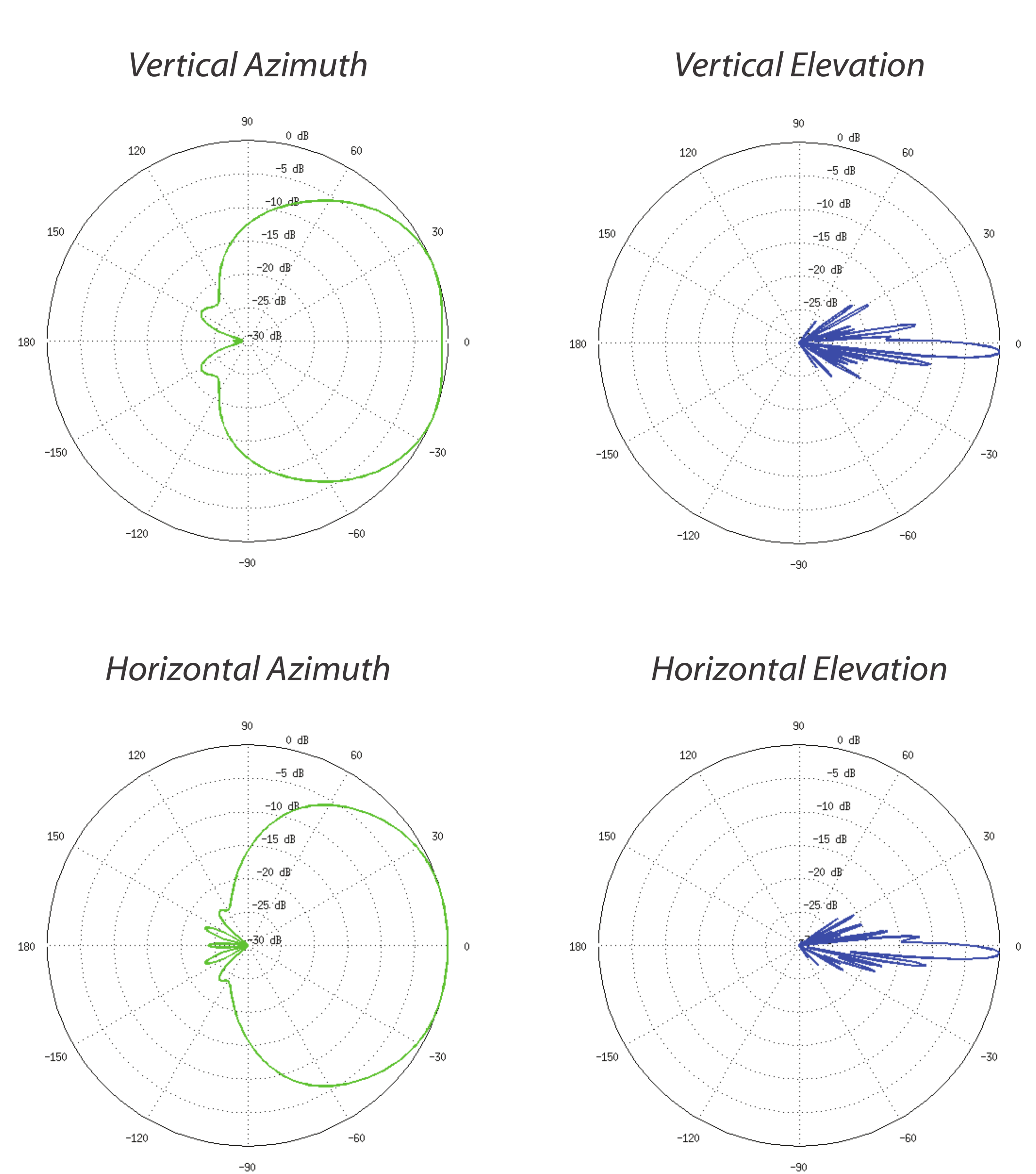}\\[2pt]
{\footnotesize (a) Base sector, AM-5G19-120}
\end{minipage}\hfill
\begin{minipage}[t]{0.48\textwidth}
\centering
\includegraphics[trim=0 0 0 8pt,clip,height=2.85in]{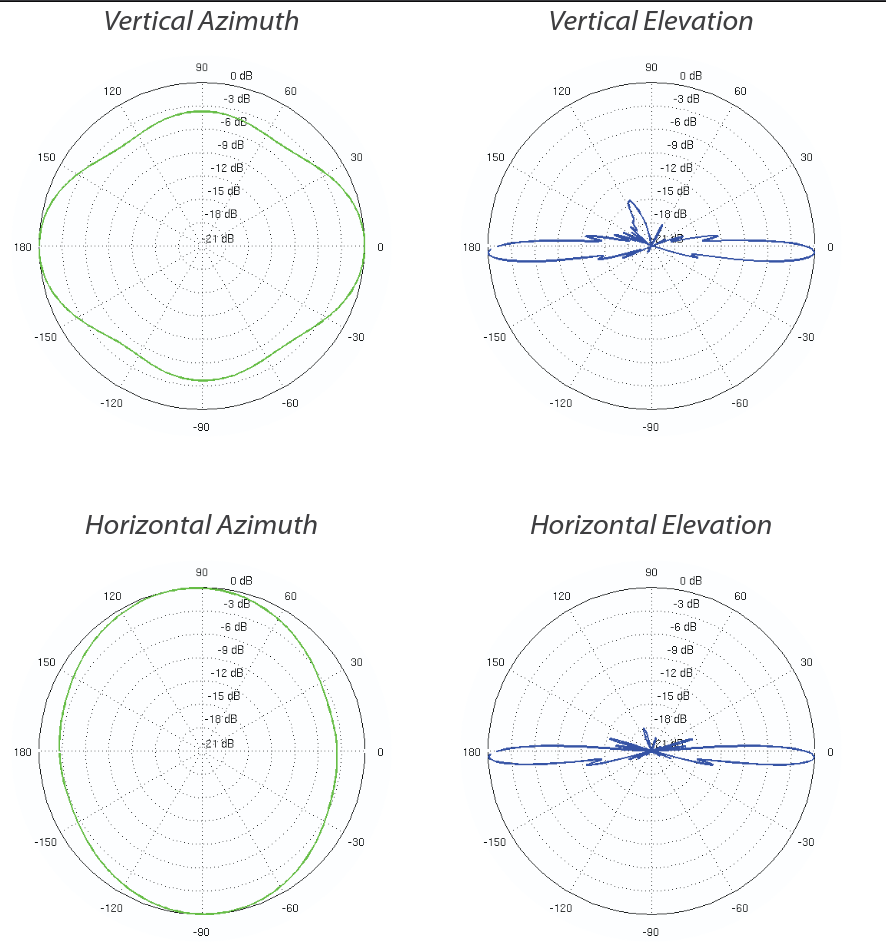}\\[2pt]
{\footnotesize (b) Rover omni, AMO-5G13}
\end{minipage}
\caption{\textbf{Manufacturer radiation patterns of the Rocket link antennas:} Patterns are normalized to the peak, so the outer ring is 0\,dB. Each panel gives the azimuth and the elevation cut for the vertically and horizontally polarized ports. In the elevation cuts, 0$^\circ$ is the horizon and 90$^\circ$ points straight up. Panel (a) is taken from~\cite{ubiquiti_sector} and panel (b) from the manufacturer's AMO-5G13 pattern plot. Both antennas have \ElecDowntilt$^\circ$ of electrical downtilt.}
\label{fig:datasheet_patterns}
\end{figure*}

The rover PC encodes each USB camera to H.264 with x264 and serves it through MediaMTX~\cite{mediamtx}. The operator views each stream over WebRTC at 720p and \Fps\,fps. At the base, an SBG Ellipse-D streams raw GNSS observations. RTKLIB's \texttt{str2str}~\cite{takasu2009rtklib} converts them to RTCM~3 MSM7 messages~\cite{rtcm10403}. An NTRIP caster on the base laptop~\cite{rtcm10410} serves them to a u-blox ZED-F9P receiver on the rover. Both streams cross the backhaul. An EdgeRouter~X~\cite{ubiquiti_erx} at each end connects to the Rocket bridge and to the Bullet bridge (Section~\ref{sec:babel}).

The base laptop writes every MAVLink command to both command radios at the same time. Kachhim, the rover's controller, therefore receives two copies of each command. It prefers the P900. SATELLINE frames carry an added cost. Kachhim executes them only after about \CmdWindow\,ms without a valid P900 frame. It returns to the P900 as soon as a valid frame arrives. The backup stream is already running, so a switch waits only for the next SATELLINE frame. This wait has not been measured.

If both radios fall silent, Kachhim's firmware stops the rover within \CmdStopTimeout\,s. In that time the rover covers about \StopDistMedian\,m at its median drive speed and \StopDistMax\,m at its maximum (Table~\ref{tab:positioning}). A firmware arm and disarm switch is the emergency stop. Both mechanisms run on Kachhim, so neither is independent of the controller. Some rover competitions require that independence~\cite{circ2026safety}.

The base laptop, Kachhim, the two routers and the rover power bus remain single points of failure. The stack duplicates radios and paths but not computers.

\section{Backhaul Link Selection with Babel}
\label{sec:babel}
The base LAN and the rover LAN are joined by two bridged subnets, one per radio link. Choosing a link is therefore a routing decision. Both routers run babeld~\cite{babeld_man}, and we checked the behavior below against its source code.

Each router sends a Hello on each link every \HelloInterval\,s. It keeps a history of which of its neighbor's last \HelloHistory{} Hellos arrived. A Hello counts as missed \HelloMissAfter\,s after the previous one. The receive cost is
\begin{equation}
  c_{\mathrm{rx}} = \CostClean \cdot 2^{15}/(s+1),
  \label{eq:rxcost}
\end{equation}
where $s$ is the history read as a binary number. The two newest bits are moved down to weight $2^{13}$, so the three newest Hellos each carry a quarter of the weight. Neighbors exchange their receive costs, and the link cost is
\begin{equation}
  C = c_{\mathrm{tx}}\, c_{\mathrm{rx}} / \CostClean + P_{\mathrm{rtt}}.
  \label{eq:cost}
\end{equation}
The delay penalty $P_{\mathrm{rtt}}$ rises linearly from zero at \RttMin\,ms of round-trip time to \RttPenaltyMax{} at \RttMax\,ms~\cite{rfc9616}. A clean link costs \CostClean. Losing the newest Hello raises the cost to \CostOneLost, the two newest to \CostTwoLost{} and the three newest to \CostThreeLost. The cost becomes infinite after \HellosToInfinity{} lost Hellos, or when a port loses carrier with link detection on.

A route's metric is the link cost plus the neighbor's advertised metric and an input filter offset. The advertised metric is zero here. The daemon switches at once if the installed route becomes infinite. Otherwise it switches only if the candidate's metric is no higher and its smoothed metric is strictly lower. The smoothing has a half-life of \SmoothHalfLife\,s, and ties never switch. Each router installs exactly one route per destination and chooses only for its own outbound traffic.

Our configuration treats both links as wireless and sends timestamped Hellos for the delay penalty. Both also use link detection. Routes go to a separate table, because EdgeOS deleted fast-changing routes in the main table. The only preference is an offset of +\Offset{} on routes over the Bullet. A clean Bullet route therefore costs \OffsetThreshold. The Rocket keeps the traffic while its cost stays at or below \OffsetThreshold, which is about two lost Hellos. It loses the traffic when its smoothed metric exceeds \OffsetThreshold{} or when its routes are withdrawn.

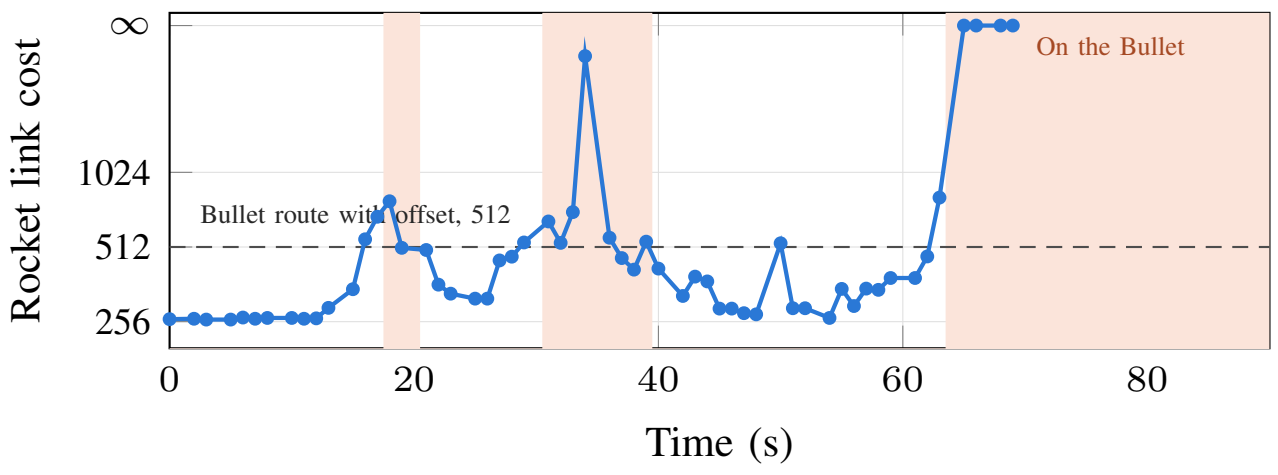
\begin{figure}[!tb]
\centering
\begin{tikzpicture}
\begin{semilogyaxis}[
  width=\columnwidth, height=3.8cm,
  xlabel={Time (s)}, ylabel={Rocket link cost},
  xmin=0, xmax=90, ymin=200, ymax=4500,
  ytick={256,512,1024,4000}, yticklabels={\CostClean,\CostTwoLost,\CostThreeLost,$\infty$},
  grid=major, grid style={line width=0.2pt, draw=gray!25},
  tick label style={font=\scriptsize}, label style={font=\footnotesize},
]
\addplot[draw=none, fill=bulletc!18, const plot mark mid] table[x=s, y expr={200 + 4300*\thisrow{bullet}}] {data/babel_episode_route.dat} \closedcycle;
\addplot[gray!60!black, densely dashed, domain=0:90, samples=2] {\OffsetThreshold};
\node[font=\tiny, gray!30!black, anchor=south west] at (axis cs:1,520) {Bullet route with offset, \OffsetThreshold};
\addplot[rocketc, thick, mark=*, mark size=0.9pt] table[x=s, y=cost] {data/babel_episode_cost.dat};
\node[font=\tiny, bulletc!70!black, anchor=north] at (axis cs:77,4300) {On the Bullet};
\end{semilogyaxis}
\end{tikzpicture}
\caption{\textbf{Rocket link cost at the range limit:} Cost at the base router as the rover passed \EdgeFrom--\EdgeTo\,m under load. Shading marks the seconds in which the base router routed over the Bullet. Traffic left the Rocket when the cost crossed \OffsetThreshold. It flapped while the cost hovered near that value and stayed on the Bullet once no Rocket Hellos were heard.}
\label{fig:babel_episode}
\end{figure}

Without the offset, any lost Hello moved traffic to the Bullet. Because ties never switch, the traffic then stayed there. It also flapped. Under our own traffic the routes changed every \FlapLo--\FlapHi\,s. Fig.~\ref{fig:babel_episode} shows the offset rule at the Rocket's range limit.

With one route per destination, the backhaul is active and standby. No flow uses both links, and the Bullet carries only Hellos while the Rocket works. Babel can still share load by destination, because an input filter can match a prefix as well as an interface. Giving the offset only to some prefixes would route those prefixes over the Bullet. The RTK stream or one camera host are examples. Each such prefix would keep the other link as its failover. Source-specific routing~\cite{boutier2015source,rfc9079} extends this to source addresses. We have not tested either option.

\section{Method}
\label{sec:method}
\begin{figure*}[t]
\centering
\includegraphics[width=\textwidth]{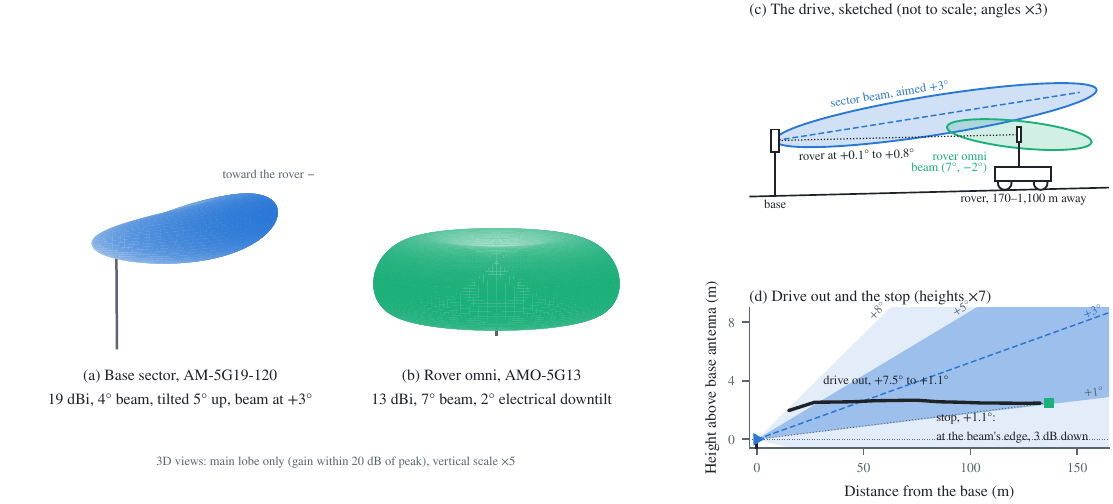}
\caption{\textbf{Antenna patterns and the rover's place in them:} (a)~Base sector and (b)~rover omni, modeled from their datasheet beamwidths with a parabolic main lobe. Only the main lobe is drawn and the vertical scale is stretched \AntVexag{} times. The sector is tilted \SectorMechTilt$^\circ$ up, so with its \ElecDowntilt$^\circ$ electrical downtilt its beam points \SectorBeam$^\circ$ up. (c)~Sketch of the drive, not to scale and with angles drawn \AntSketchEx{} times larger. The sector's beam passes over the rover, which the base saw at +\ThetaMin$^\circ$ to +\ThetaMax$^\circ$. The rover therefore sat at the beam's lower edge, with an estimated \PatternLossMin--\PatternLossMax\,dB of pattern loss. The base sat inside the rover omni's \OmniEl$^\circ$ beam, which its electrical downtilt aims \ElecDowntilt$^\circ$ down. (d)~Scaled side view of the drive out and the cable-repair stop, with heights stretched by the factor shown. The dark wedge shows where the sector's gain is within half power of its peak, and the light wedge where it is within \PatternFloor\,dB.}
\label{fig:antennas}
\end{figure*}

\subsection{Command Failover Test}
\label{sec:met_command}
The P900 was powered off abruptly while commands were flowing, in \CtrlTrials{} trials. Kachhim's window for the primary was about \CmdWindow\,ms.

\subsection{Bench Backhaul Tests}
\label{sec:met_bench}
Both radio pairs stood on the bench in line of sight. A script applied Linux netem~\cite{hemminger2005netem} to the Rocket ports of both routers. It lost or delayed packets while the radios stayed associated, as a weak or obstructed link would. Each run stepped through phases of \PhaseLo--\PhaseHi\,s with rising impairment. Distance runs combined loss with added delay, and noise runs added loss alone. Each run then blocked all traffic and finally removed the impairment. In a separate trial the base Rocket was powered off while it carried traffic.

An ICMP echo every \EchoInterval\,ms from the base laptop to the rover PC measured each outage. The outage is the length of the first unbroken run of lost echoes. Each router's installed route was read every second. The +\Offset{} offset was added between two series of runs, so each impairment was tested with and without it. Runs with a ping accounting error are excluded. So are two blocked runs in which the script's control connection stalled over the impaired link. During the last blocked trial, a probe received one video stream over UDP and over TCP and logged every frame. As a reference, iperf3 measured single-stream TCP throughput with all traffic on the Rocket.

\subsection{Field Drive}
\label{sec:met_drive}

The drive took place on open ground in the Purbachal helipad area of Dhaka, at \SiteLat$^\circ$N and \SiteLon$^\circ$E. The route ran east from the base antennas. It was in line of sight to about \LosTo\,m. Beyond that point the route followed a road lined with trees on both sides, which blocked the line of sight to the base antennas. This ground differs from the competition site. URC runs in the desert around the Mars Desert Research Station, on rocky and uneven ground that includes a canyon~\cite{urc_requirements,urc_delivery_2025}. Our route had no such relief, and its only obstruction was trees.

The rover stood at \SetupAt\,m during setup, which is not analyzed. It then drove out to \StopFrom\,m and stood at \StopFrom--\StopTo\,m for \StopMinutes\,min while a torn motor driver cable was replaced. The radios and loggers kept running, so this stop is the static reference. The rover then drove east from \DriveStart\,m to \DriveMax\,m in \DriveMinutes\,min.

One iperf3 round ran at \FieldTcpFrom--\FieldTcpTo\,m. Between \LoadFrom{} and \LoadTo\,m, iperf3 rounds of \LoadRate\,Mb/s ran every \LoadEvery\,min. The rest of the drive carried only the video, the corrections and the loggers.

Table~\ref{tab:logged} lists what was logged and how often. Rover logs were re-timed to GNSS time before they were joined with base logs. After the Rocket station became unreachable near \RocketLost\,m, the radio poller backed off to one poll every \BackoffLo--\BackoffHi\,min. Radio-reported signal and capacity are therefore dense only to \DenseTo\,m.

\begin{table}[t]
\centering
\caption{Quantities Logged on the Drive}
\label{tab:logged}
\footnotesize
\setlength{\tabcolsep}{3pt}
\begin{tabularx}{\columnwidth}{@{}YL{2.5cm}L{1.5cm}@{}}
\toprule
\textbf{Quantity} & \textbf{Source} & \textbf{Rate} \\
\midrule
Signal, capacity, MCS and temperature of the four video radios & Base logger & Every \PollRadios\,s \\
Echo loss and round-trip time & Base laptop to rover PC & Every \PollPing\,s \\
Installed routes and link costs & Babel daemon on both routers & About every second \\
Position, solution and accuracy & ZED-F9P on the rover & \RoverEpochRate\,Hz \\
RTCM frames and correction age & Rover PC & Every frame \\
One \LogVideoW$\times$\LogVideoH{} stream over RTSP and TCP & Base logger & Every frame \\
\bottomrule
\end{tabularx}
\end{table}

\subsection{Analysis}
\label{sec:met_analysis}

Distance is horizontal, from the rover's RTK position to the base's \BaseAvgSeconds\,s averaged position. Signal, capacity and service are summarized as medians over \BinM\,m distance bins, and bins with fewer than eight samples are dropped.

For each link we fit the log-distance model
\begin{equation}
  P(d) = P_{\FitRef} - 10\, n \log_{10}\!\left(d / \FitRef\,\mathrm{m}\right)
  \label{eq:pathloss}
\end{equation}
by least squares to the bin medians of the drive beyond \FitFrom\,m. Here $P_{\FitRef}$ is the fitted power at \FitRef\,m and $n$ is the path loss exponent. We report $n$ with its \CiLevel\% confidence interval from the $t$ distribution.

The shortfall is the gap between the free-space prediction and each measured Rocket sample at the rover end. The prediction uses the configured transmit power, both datasheet antenna gains and \FsMiscLoss\,dB of other loss.

The rover's elevation $\theta$ seen from the base follows from its RTK height above the base antenna and its distance. Each antenna's vertical pattern is modeled as a parabolic main lobe,
\begin{equation}
  A(\theta) = -\min\!\left( \PatternK \left( \frac{\theta - \theta_{\mathrm{tilt}}}{\theta_{\mathrm{3dB}}} \right)^{2},\; \PatternFloor \right)\ \mathrm{dB},
  \label{eq:pattern}
\end{equation}
with the beam elevation $\theta_{\mathrm{tilt}}$ and the vertical beamwidth $\theta_{\mathrm{3dB}}$ from Table~\ref{tab:links}. The pattern loss of the link is the sum of the sector's loss at $\theta$ and the omni's loss at $-\theta$.

Echo loss and round-trip time come from the echoes every \PollPing\,s. Correction delivery is the share of the nominal \RtcmNominal\,B/s RTCM stream that arrived at the rover over each bin. Age counts fresh corrections, those at most \AgeLimit\,s old. A second counts as smooth video when at least \SmoothFrames{} of its \Fps{} frames arrived.

Finally, for each capacity sample with both links up, we compare the capacity of the link in use with that of the other link.

\section{Results}
\label{sec:results}

\subsection{Command Failover}
\label{sec:res_failover}
All \CtrlTrials{} switchovers succeeded. The mean was \CtrlSwitchMean\,ms with a \CiLevel\% confidence interval of \CtrlSwitchCILo--\CtrlSwitchCIHi\,ms. Single trials ranged from \CtrlSwitchMin{} to \CtrlSwitchMax\,ms. The mean is consistent with the window of about \CmdWindow\,ms.

\subsection{Backhaul Failover on the Bench}
\label{sec:res_backhaul}

A dead primary was detected faster than a silent one. When the base Rocket lost power, link detection withdrew its routes at once. Traffic moved to the Bullet with a \PowerOffOutage\,s outage. With all packets lost and the radios still associated, Babel had to notice the missing Hellos instead. The \BlockTrials{} valid trials gave outages of \BlockOutageMin--\BlockOutageMax\,s. This is consistent with the cost rule of Section~\ref{sec:babel}, which needs several lost Hellos to overcome the offset.

Traffic returned to the Rocket \ReturnMin--\ReturnMax\,s after each impairment ended, in all \ReturnTrials{} recovery phases. The smoothing likely sets this delay, since the Bullet route is kept until the Rocket's smoothed metric is strictly lower.

What the operator saw depended on the transport protocol. Over UDP the longest video freeze in a blocked trial was \FreezeUdp\,s, close to the outage. Over TCP, retransmission backoff stretched it to \FreezeTcp\,s.

Babel tolerated a lossy primary only with the offset in place (Table~\ref{tab:ablation}). Without it, \NoOffsetLoss\% loss with \NoOffsetDelay\,ms of delay each way moved traffic after \NoOffsetAfter\,s, and the traffic never came back. With the offset, Babel stayed on a Rocket that lost \OffsetHoldsLoss\% of packets each way with \OffsetHoldsDelay\,ms of added delay. It moved at \OffsetMovesLoss\% loss with \OffsetMovesDelay\,ms of delay, or at \NoiseMovesLoss\% loss without delay. Round-trip loss reached about a third before the switch. The offset thus trades fast failover for stability. The right setting depends on how much loss the video and the corrections can absorb.

\begin{table}[t]
\centering
\caption{Babel Under Injected Loss on the Rocket}
\label{tab:ablation}
\footnotesize
\setlength{\tabcolsep}{3.4pt}
\begin{tabular}{@{}lccl@{}}
\toprule
\textbf{Impairment, each way} & \textbf{RT loss} & \textbf{RTT (ms)} & \textbf{Babel} \\
\midrule
\TabAblation
\bottomrule
\end{tabular}

\vspace{2pt}
{\footnotesize RT loss and RTT: loss and median round-trip time of the \EchoInterval\,ms echoes over each phase. The bursty rows used correlated netem loss, which dropped far fewer packets than its nominal rate. Without the offset, traffic never returned to the Rocket.}
\end{table}

\subsection{Signal and Range Along the Drive}
\label{sec:res_signal}

The drive ran on a degraded Rocket. At the stop, the base Rocket heard \ChainZeroDbm\,dBm on chain~0 against \ChainOneDbm\,dBm on chain~1, so one receive chain was effectively dead. The link ran at MCS~\RocketMcsMin{} to~\RocketMcsMax, with one spatial stream in at least \RocketOneStreamPct\% of reports. It reported \CapStopRocket\,Mb/s of downlink capacity against \CapStopBullet\,Mb/s for the Bullet at MCS~\BulletMcs. Both Rockets and the rover Bullet also linked at only \EthRocket\,Mb/s on Ethernet.

The rover Rocket's board warmed in the sun from \TempRoverFirst{} to \TempRoverMax\,$^\circ$C. It peaked at \TempRoverMaxAt\,m, shortly before the link was lost. The base Rocket stayed at \TempBaseFirst--\TempBaseMax\,$^\circ$C. Load rounds began in the same period, so we do not attribute the loss to heat.

\begin{figure}[!tb]
\centering
\begin{tikzpicture}
\begin{groupplot}[
  group style={group size=1 by 3, vertical sep=9pt, xticklabels at=edge bottom, xlabels at=edge bottom},
  width=\columnwidth, height=4.0cm, xmin=0, xmax=1150, xtick={0,200,400,600,800,1000},
  grid=both, grid style={line width=0.2pt, draw=gray!25}, minor grid style={draw=none},
  tick label style={font=\scriptsize}, label style={font=\footnotesize},
  y tick label style={text width=1.9em, align=right},   %
  xlabel={Distance from base (m)},
]
\nextgroupplot[ylabel={Height above MSL (m)}, ymin=4, ymax=12, ytick={4,6,8,10,12}]
\fill[gray!14] (axis cs:\DipOneFrom,4) rectangle (axis cs:\DipOneTo,12);
\fill[gray!14] (axis cs:\DipTwoFrom,4) rectangle (axis cs:\DipTwoTo,12);
\node[font=\tiny, gray!40!black, anchor=north] at (axis cs:\DipOneMid,11.9) {Dip, $-$\DipOneDepth\,m};
\node[font=\tiny, gray!40!black, anchor=north] at (axis cs:\DipTwoMid,11.9) {Dip, $-$\DipTwoDepth\,m};
\addplot[gray!70!black, densely dashed, domain=0:1150, samples=2] {\BaseMslPlot};
\addplot[only marks, mark=triangle*, mark size=2.4pt, gray!40!black] coordinates {(0,\BaseMslPlot)};
\node[font=\tiny, gray!30!black, anchor=south west] at (axis cs:8,\BaseMslPlot) {Base antenna (SBG Ellipse-D), \BaseMsl\,m};
\addplot[black!85, thick] table[x=d, y=msl] {data/drive_elev.dat};
\node[font=\tiny, black!85, anchor=south] at (axis cs:140,9.2) {Rover antenna (RTK)};
\draw[gray, dotted, thick] (axis cs:\RtkToPlot,4) -- (axis cs:\RtkToPlot,12);
\node[font=\tiny, gray!40!black, anchor=north east, align=right] at (axis cs:1110,7.3) {No RTK\\beyond\\\RtkTo\,m};
\node[font=\scriptsize, anchor=north west] at (rel axis cs:0.005,0.99) {(a)};
\nextgroupplot[ylabel={RSSI at rover (dBm)}, ymin=-95, ymax=-40]
\fill[gray!14] (axis cs:\DipOneFrom,-95) rectangle (axis cs:\DipOneTo,-40);
\fill[gray!14] (axis cs:\DipTwoFrom,-95) rectangle (axis cs:\DipTwoTo,-40);
\addplot[rocketc, thick, mark=*, mark size=1.4pt, error bars/.cd, y dir=both, y explicit, error bar style={rocketc, line width=0.5pt}]
  table[x=d, y=med, y error minus=lo, y error plus=hi] {data/drive_rssi_rocket.dat};
\addplot[bulletc, thick, densely dashed, mark=square*, mark size=1.3pt, mark options={solid}, error bars/.cd, y dir=both, y explicit, error bar style={bulletc, solid, line width=0.5pt}]
  table[x=d, y=med, y error minus=lo, y error plus=hi] {data/drive_rssi_bullet.dat};
\node[font=\tiny, rocketc!80!black, anchor=north] at (axis cs:60,-74) {Rocket};
\node[font=\tiny, bulletc!80!black, anchor=south] at (axis cs:110,-53.5) {Bullet};
\draw[gray, dotted, thick] (axis cs:\RocketLost,-95) -- (axis cs:\RocketLost,-40);
\node[font=\tiny, gray!40!black, anchor=north west] at (axis cs:780,-42) {Rocket lost, \RocketLost\,m};
\node[font=\scriptsize, anchor=north east] at (rel axis cs:0.995,0.18) {(b)};
\nextgroupplot[ylabel={Capacity (Mb/s)}, ymin=0, ymax=70,
  legend cell align=left, legend columns=2,
  legend style={font=\tiny, draw=none, at={(0.5,-0.36)}, anchor=north, /tikz/every even column/.append style={column sep=6pt}}]
\fill[gray!14] (axis cs:\DipOneFrom,0) rectangle (axis cs:\DipOneTo,70);
\fill[gray!14] (axis cs:\DipTwoFrom,0) rectangle (axis cs:\DipTwoTo,70);
\addplot[rocketc, thick, mark=*, mark size=1.4pt] table[x=d, y=rocket] {data/capacity_walk.dat};
\addlegendentry{Rocket, \RocketGHz\,GHz (carried the traffic)}
\addplot[bulletc, thick, densely dashed, mark=square*, mark size=1.3pt, mark options={solid}] table[x=d, y=bullet] {data/capacity_walk.dat};
\addlegendentry{Bullet, \BandTwoFour\,GHz (idle backup)}
\draw[gray, dotted, thick] (axis cs:\RocketLost,0) -- (axis cs:\RocketLost,70);
\node[font=\scriptsize, anchor=north east] at (rel axis cs:0.995,0.99) {(c)};
\end{groupplot}
\end{tikzpicture}
\caption{\textbf{Rover antenna height, signal and capacity along the drive:} (a)~Rover antenna height above MSL from its RTK float solution in \HeightBinM\,m medians, and the base antenna's height as a dashed line. (b)~Signal at the rover end of each link, in \BinM\,m medians with interquartile bars. (c)~Radio-reported downlink capacity in \BinM\,m medians. The base router used the Rocket in every sample from \DriveStart{} to \CapRocketTo\,m. Shading marks the two antenna dips. Beyond \DenseTo\,m the poller was in back-off.}
\label{fig:walk_rssi}
\end{figure}
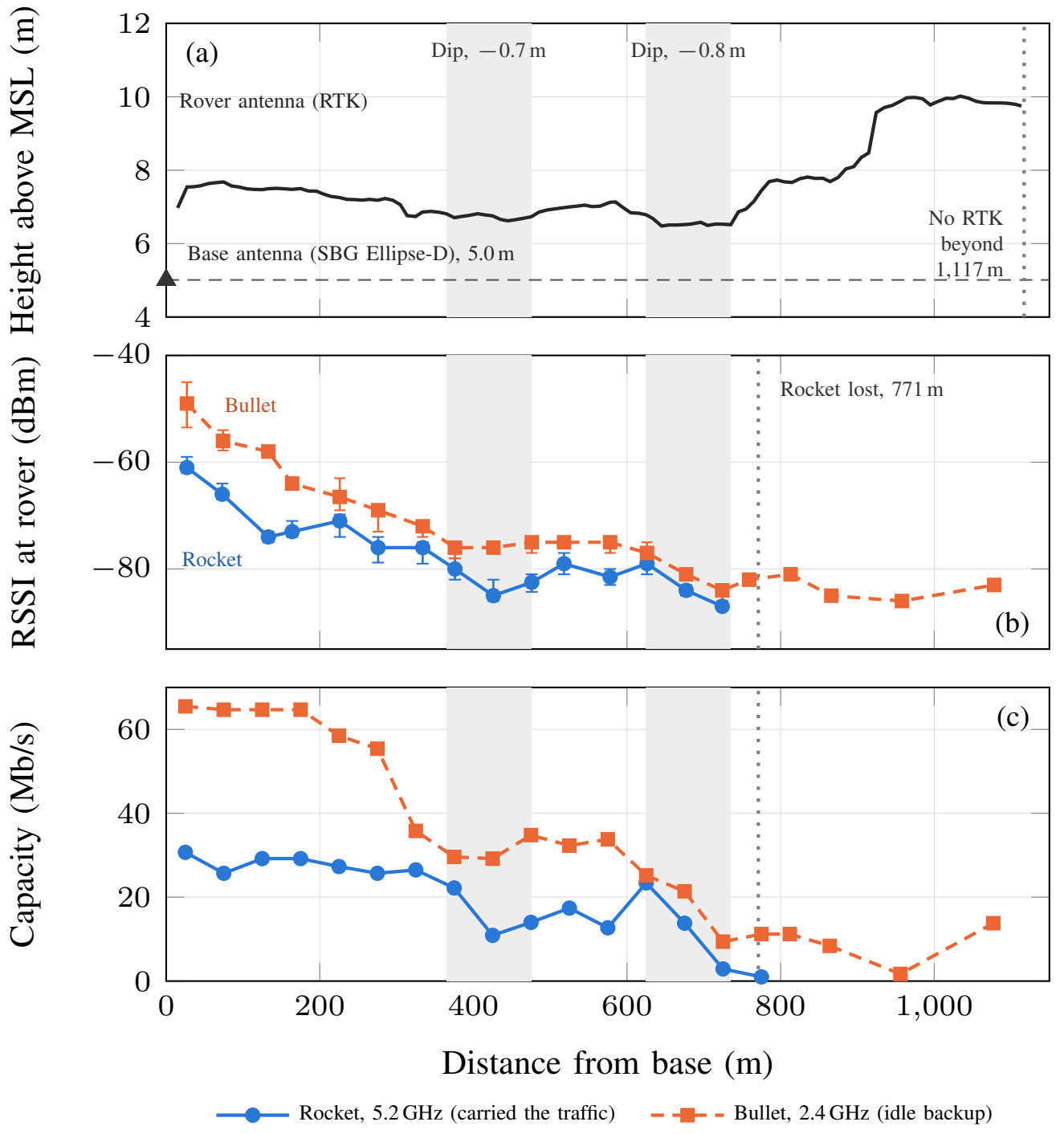

Signal fell with distance at close to the free-space slope on both links (Fig.~\ref{fig:walk_rssi}b). At the rover end the fits give exponents of \PlExpRocket{} for the Rocket and \PlExpBullet{} for the Bullet (Table~\ref{tab:pathloss}). The intervals are wide because each fit rests on only twelve or thirteen bins. The fits cover the line-of-sight section and the first part of the obstructed one.

\begin{table}[t]
\centering
\caption{Log-Distance Fits to the Drive Medians}
\label{tab:pathloss}
\footnotesize
\setlength{\tabcolsep}{3pt}
\begin{tabular}{@{}lccccc@{}}
\toprule
\textbf{Link, end} & \textbf{$n$ (\CiLevel\% CI)} & \textbf{$P_{\FitRef}$ (dBm)} & \textbf{$\sigma$ (dB)} & \textbf{$R^2$} & \textbf{Bins} \\
\midrule
\TabPathloss
\bottomrule
\end{tabular}

\vspace{2pt}
{\footnotesize Fits of~\eqref{eq:pathloss} to the \BinM\,m medians beyond \FitFrom\,m. $\sigma$: residual standard deviation. The base Rocket's dead chain~0 is left out.}
\end{table}

The Rocket's signal sat \ShortfallMin--\ShortfallMax\,dB below free space. It was as large within line of sight as beyond it, at \ShortfallLosMin--\ShortfallLosMax\,dB against \ShortfallObsMin--\ShortfallObsMax\,dB. Trees therefore do not explain it.

Antenna aiming explains only a small part of the shortfall. Seen from the base, the rover stayed \ThetaMin$^\circ$ to \ThetaMax$^\circ$ above the horizontal. The sector's \SectorEl$^\circ$ beam pointed \SectorBeam$^\circ$ up, so its lower half-power edge was at \SectorLowEdge$^\circ$ (Fig.~\ref{fig:antennas}). For the whole drive, then, the rover sat at or just below that edge. Yet the model of~\eqref{eq:pattern} predicts only \PatternLossMin--\PatternLossMax\,dB of pattern loss for the two antennas together.

The rest of the shortfall is unexplained. The rover omni is specified only for \OmniBandLo--\OmniBandHi\,GHz~\cite{ubiquiti_amo5g13}. Its gain at \RocketGHz\,GHz is therefore unknown and may account for part of the gap. The base Rocket's cabling and alignment are the next things to check.

The rover antenna dipped \DipOneDepth{} and \DipTwoDepth\,m below its earlier level (Fig.~\ref{fig:walk_rssi}a). The Rocket felt it. In the first dip its median fell to \DipOneRocket\,dBm, while the Bullet's held at \DipOneBullet\,dBm. On the drive out, the rover passed down through the sector's beam. Its elevation fell from \ThetaOutStart$^\circ$ at \ThetaOutStartAt\,m to \ThetaOutEnd$^\circ$ at \ThetaOutEndAt\,m (Fig.~\ref{fig:antennas}d).

Rocket Hellos stopped at \RocketLost\,m, which is \RocketLostBeyondLos\,m into the obstructed section and during a load round. Apart from a few seconds near \RocketBlip\,m they did not return. Bullet Hellos stopped at \BulletLost\,m. The last echo reply arrived at \LastPing\,m and the last RTCM frame at \LastRtcm\,m.

An earlier configuration of the video link was also used at URC, but no data were logged there. In those runs, the operators found the video link usable to about \CompRangeLo--\CompRangeHi\,m, depending on the terrain and the task. This is an operator estimate and not a measurement. It brackets the \RocketLost\,m that the Rocket reached on the drive in Dhaka.

\subsection{Service Along the Drive}
\label{sec:res_service}
Without load, the backhaul served the rover well out to \LoadFrom\,m. It lost at most \PingLossIdleMax\% of echoes and gave smooth video in at least \SmoothIdleMin\% of seconds (Table~\ref{tab:walk_service}). Under the load rounds between \LoadFrom{} and \LoadTo\,m, smooth video fell to \SmoothLoad\% of seconds. Correction delivery fell to \RtcmLoad\% of the nominal bytes, and \AgeLoad\% of epochs had a fresh correction. The same band also holds the Rocket's failure at \RocketLost\,m, so the effects of load and range cannot be separated there. In the band where the Bullet failed, echo loss rose to \PingLossEdge\%.

The logged video stream itself was light. At the stop it averaged \VideoRateStop\,Mb/s, with a \PctlLevel th percentile of \VideoRatePctl\,Mb/s.

\begin{table}[t]
\centering
\caption{Service Along the Drive}
\label{tab:walk_service}
\footnotesize
\setlength{\tabcolsep}{2.6pt}
\begin{tabular}{@{}lcccccc@{}}
\toprule
\textbf{Distance} & \textbf{Load} & \textbf{Ping} & \textbf{RTT} & \textbf{RTCM} & \textbf{Age} & \textbf{Smooth} \\
\textbf{(m)} & & \textbf{loss} & \textbf{(ms)} & \textbf{bytes} & \textbf{$\leq$\AgeLimit\,s} & \textbf{video} \\
\midrule
\TabService
\bottomrule
\end{tabular}

\vspace{2pt}
{\footnotesize RTCM bytes: share of the nominal \RtcmNominal\,B/s that arrived. Age: rover epochs with a correction at most \AgeLimit\,s old. Smooth video: seconds with at least \SmoothFrames{} of \Fps{} frames.}
\end{table}

\subsection{Positioning}
\label{sec:res_rtk}
RTK accuracy held beyond the Rocket's range (Table~\ref{tab:positioning}). The receiver reported \StaticH\,cm horizontal accuracy at the stop. It reported \NearH\,cm within line of sight and \FarH\,cm beyond it, where the Bullet carried most corrections. These are receiver estimates and therefore optimistic. The rare fixed epochs jumped about one meter against the float track, so we treat the position as good to decimeters.

The base position is a \BaseAvgSeconds\,s average of the Ellipse-D's SBAS solution. Its absolute error, about \BaseAbs\,m, shifts every rover position equally. Correction latency was \RtcmLatency\,ms at the median and \RtcmLatencyPctl\,ms at the \PctlLevel th percentile, while the echo round-trip time over that span was \RtcmRtt\,ms. Latency therefore arises at the base.

\begin{table}[!tb]
\centering
\caption{Positioning and Rover Motion on the Drive}
\label{tab:positioning}
\footnotesize
\begin{tabularx}{\columnwidth}{@{}>{\raggedright\arraybackslash}X>{\raggedright\arraybackslash}p{3.2cm}@{}}
\toprule
\textbf{Quantity} & \textbf{Value} \\
\midrule
\multicolumn{2}{@{}l}{\emph{Base station, SBG Ellipse-D, \BaseAvgSeconds\,s SBAS average}} \\
Scatter, north / east / up (1$\sigma$) & \BaseStdN{} / \BaseStdE{} / \BaseStdU\,cm \\
Absolute accuracy & About \BaseAbs\,m \\
GNSS antenna height & \BaseMsl\,m above MSL \\
\midrule
\multicolumn{2}{@{}l}{\emph{Rover, ZED-F9P, RTK float, receiver-reported}} \\
Static at the stop, horizontal / vertical & \StaticH{} / \StaticV\,cm \\
Drive \NearFrom--\LosTo\,m, horizontal / vertical & \NearH{} / \NearV\,cm \\
Drive \LosTo--\FarTo\,m, horizontal / vertical & \FarH{} / \FarV\,cm \\
Drive, \PctlLevel th percentile, horizontal / vertical & \PctlH{} / \PctlV\,cm \\
Solution share on the drive & Float \ShareFloat\%, none \ShareNone\%, fixed \ShareFixed\% \\
\midrule
\multicolumn{2}{@{}l}{\emph{Rover motion on the drive}} \\
Speed, median / \PctlLevel th percentile / max & \SpeedMedian{} / \SpeedPctl{} / \SpeedMax\,m/s \\
Drive, \DriveStart{} to \DriveMax\,m & \DriveMinutes\,min \\
\bottomrule
\end{tabularx}

\vspace{2pt}
{\footnotesize Rover values are medians of epochs with an RTK float solution unless stated. Speed counts epochs moving faster than \MovingSpeed\,m/s.}
\end{table}

\subsection{Route Choice and Switching}
\label{sec:res_route}

Babel kept traffic on the lower-capacity link for most of the drive. The Bullet reported more capacity than the Rocket in every \BinM\,m bin to \DenseTo\,m (Fig.~\ref{fig:walk_rssi}c). After setup, the base router used the Rocket in \OnRocketPct\% of the \BothUpSamples{} samples with both links up, as the offset intends. In \OnLowerPct\% of those samples, the link in use reported at least \LowerMargin\% less capacity than the other link. The mean was \UsedMean{} against \BestMean\,Mb/s. Babel judged the Rocket healthy because it still delivered its Hellos. Its dead receive chain cut capacity but not Hello delivery.

The two routers used different links for \MismatchSeconds\,s. This is \MismatchPct\% of the \MismatchWindow\,min from the start of the drive until both links were lost. Counting both routers, they switched \SwitchesTotal{} times in \EpisodesWord{} episodes. The first episode had \EpOneSwitches{} switches in \EpOneSpan\,s when iperf3 saturated the Rocket at \EpOneAt\,m. It left a \EpOneRtcmGap\,s gap in the corrections. The second had \EpTwoSwitches{} switches in \EpTwoSpan\,s at the range limit under load. It caused a \EpTwoRtcmGap\,s correction gap and a \EpTwoVideo\,s video outage. The third had \EpThreeSwitches{} switches in \EpThreeSpan\,s near \EpThreeAt\,m, where the Rocket was briefly heard again.

Single-stream TCP at \FieldTcpFrom--\FieldTcpTo\,m reached \FieldTcpDown\,Mb/s down and \FieldTcpUp\,Mb/s up. On the bench, with all traffic on a healthy Rocket, the same test reached \BenchTcpDownMin--\BenchTcpDownMax\,Mb/s down and \BenchTcpUp\,Mb/s up. The gap likely reflects the dead receive chain and the \EthRocket\,Mb/s ports as well as the distance.

\section{Discussion}
\label{sec:discussion}
The drive reversed the assumption behind the offset, that the Rocket is the better link. A hardware fault left the Rocket alive but slow, and Babel cannot see this because its cost counts only lost Hellos. A watcher that raises the Rocket's cost from its reported capacity or SNR would let the routers react. The same watcher would also catch a primary that is alive but too slow for bulk data.

Without an offset, Babel flapped under our own traffic. With it, a blocked primary cost about \BlockOutageMax\,s of outage, and a lossy primary was kept until about a third of round trips failed. A shorter Hello interval should shorten the blocked outage at the cost of more control traffic. This has not been tested.

Two of the three switching episodes came under iperf3 load. Saturating the primary raises its delay and loss, and both raise its cost. Bulk transfers should therefore be rate limited on a link that also carries video and corrections.

At the \MastHeight\,m height limit, the rover stayed within one degree of the horizontal for the whole drive beyond the stop. A sector tilted up places that whole range at the lower edge of its beam. Tilting it to the horizon would put the rover inside the main lobe and recover the few decibels that the model attributes to aiming.

End-to-end echoes revealed neither the \EthRocket\,Mb/s ports nor the dead receive chain. One unreachable radio also slowed the polling of all four. Port speeds, per-chain signal and per-device poll failures should therefore be logged from the start of every test.

The backhaul failed over within \PowerOffOutage\,s of a power loss and \BlockOutageMax\,s of a blockage. Both are shorter than ArduPilot Rover's \ArduGcs\,s ground station failsafe~\cite{ardupilot_rover_failsafe}, so a rover with this backhaul would change links before that failsafe fired. The team reports of Section~\ref{sec:related} give no switch times to compare against.

\section{Limitations}
\label{sec:limits}
All measurements were made in Dhaka. At URC an earlier configuration of the video link ran without logging. Its competition range (Section~\ref{sec:res_signal}) is therefore an operator estimate. Desert relief and the URC command trailers may change both range and multipath.

The bench test was in line of sight. The drive was in line of sight only to about \LosTo\,m, so its range results hold for one partly obstructed path.

The backhaul results come from one bench session and one drive. The drive ran on a degraded Rocket, so its range understates a healthy link.

Only one video stream was logged, over RTSP and TCP. The operator's WebRTC streams were not measured.

The command switchover rests on \CtrlTrials{} trials, and the safe stop has not been timed.

The RTK accuracy figures are receiver estimates without an independent reference.

\section{Conclusion}
Separating vision and control by radio and band let each part fail on its own. On the bench, Babel moved the backhaul to the backup within \PowerOffOutage\,s of a power loss and \BlockOutageMax\,s of a blockage. In the field, the backup extended service beyond the line-of-sight section from \RocketLost\,m to \BulletLostKm\,km, with RTK accuracy near its static value. Teams that copy this design should not judge a link by Hello loss alone. Here a degraded but alive primary kept the traffic while the backup offered twice its capacity. Next are a capacity-aware backhaul metric, a re-aimed sector and logged tests at the competition site.

\section*{Acknowledgment}
We thank Rafid Khan, former team lead, and Mollah Md Saif, co-advisor, of the BRACU Mongol-Tori Rover Team. We also thank the rest of the team and BRAC University's Department of Computer Science and Engineering.

\bibliographystyle{IEEEtran}
\bibliography{refs_fetched,refs_fetched2,refs_manual}

\end{document}